\documentclass[pre,twocolumn,floatfix,bibnotes]{revtex4}

\usepackage{here}
\usepackage{graphicx}
\usepackage{bm}

\newcommand\pictc[5]{\begin{figure}
                       \centerline{
                       \includegraphics[width=#1\columnwidth]{#3}}
                   \protect\caption{\protect\label{fig:#4} #5}
                    \end{figure}            }
\newcommand\pict[4][1.]{\pictc{#1}{!tb}{#2}{#3}{#4}}
\newcommand\rpict[1]{\ref{fig:#1}}

\newcounter{Fig}

\begin{document}

\title{Bistable diode action in left-handed periodic structures}

\author{Michael W. Feise, Ilya V. Shadrivov, and Yuri S. Kivshar}

\affiliation{Nonlinear Physics Centre and Centre for Ultra-high
bandwidth Devices for Optical Systems (CUDOS), Research School of
Physical Sciences and Engineering, Australian National University,
Canberra, ACT 0200, Australia}

\begin{abstract}
We study nonlinear transmission of an asymmetric multilayer
structure created by alternating slabs of two materials with
positive and negative refractive index. We demonstrate that such a
structure exhibits passive spatially nonreciprocal transmission of
electromagnetic waves,
the analogue of the electronic diode. We study the properties of
this left-handed diode and confirm its highly nonreciprocal and
bistable transmittance by employing direct simulations.
\end{abstract}

\maketitle

An all-optical diode is a spatially nonreciprocal device that
allows unidirectional propagation of a signal at a given
wavelength. In the ideal case, the diode transmission is 100\% in
the forward propagation, whereas it is much smaller or vanishes
for backward (opposite) propagation, yielding a unitary contrast.
The possibility of achieving highly nonreciprocal transmission
through a passive, nonlinear device is not only novel and
interesting concept, but also useful for many applications such as
optical isolation and all-optical processing. Such unidirectional
action was demonstrated for several asymmetric structures and
different nonlinear
materials~\cite{Scalora:1994-2023:JAP,Tocci:1995-2324:APL,Liang:1997-1192:APL,Gallo:1999-267:JOSB,Mujumdar:2001-929:OL,Gawith:2001-4106:APL,Gallo:2001-314:APL}.

In this paper, we discuss a novel type of spatially nonreciprocal
device based on an asymmetric multilayered structure created by
alternating slabs of two materials, conventional and left-handed (LH),
the latter is a material with both negative electric permittivity and
negative magnetic permeability which results in an effective negative
refractive index~\cite{LHM-foucs-issue:2003:OE}.  Such multilayered
structures are a special class of structures consiting of a sequence
of flat lenses that provide optical
cancellation of the conventional, or right-handed (RH), layers leading
to either enhanced or suppressed
transmission~\cite{Nefedov:2002-36611:PRE,Li:2003-83901:PRL,Feise:2004-1451:APL}.

We employ a general idea to obtain diode action by making the
structure both \emph{asymmetric} and 
\emph{nonlinear}. Also, we consider a periodic superlattice, which is
expected to possess \emph{resonant} properties, and thus enhancing
nonlinear effects. 
We consider an asymmetric superlattice consisting of three stacks
with four LH/RH double layers each [see
Figs.~\rpict{structure}(a,b)]. We assume that in the nonlinear
regime, a dielectric defect layer with Kerr-type nonlinear
response is inserted between stacks 1 and 2, as shown in
Fig.~\rpict{structure}(b). We study wave transmission in both linear
and nonlinear regimes by using the transfer-matrix method
(TMM)~\cite{Yeh:1988:OpticalWaves}, and also by employing direct
numerical simulations based on the pseudospectral time-domain
(PSTD) method \cite{Liu:1997-158:MOTL}. The transfer matrix of the
system analytically relates the electromagnetic fields on either side
of an element 
\cite{Yeh:1988:OpticalWaves}.  Complex structures can be build up
out of simple elements through multiplication of the respective matrices.
Thin nonlinear elements can be modelled, e.g., as delta-functions
\cite{Lidorikis:1998-346:PD,Feise:2004-1451:APL}.

In the PSTD method, the Maxwell equations are discretized in time
and space. The spatial derivatives are approximated using discrete
Fourier transforms and the temporal derivatives using central
differences. From this, one derives update equations for the
different fields and by iteration propagates the fields through
time. The problem of the inherent periodicity of the Fourier
transform can be removed through the use of
perfectly-matched-layer absorbing boundary-conditions
\cite{Liu:1997-158:MOTL,Berenger:1994-185:JCP}. The PSTD method
samples all field components at the same location in the unit
cell, which is advantageous when both the electric permittivity
and the magnetic permeability vary with position
\cite{Feise:2004-2955:ITAP}.  An instantaneous Kerr nonlinear
material is included by directly solving the cubic equation for
the intensity~\cite{Tran:1996-1138:OL} at each iteration.

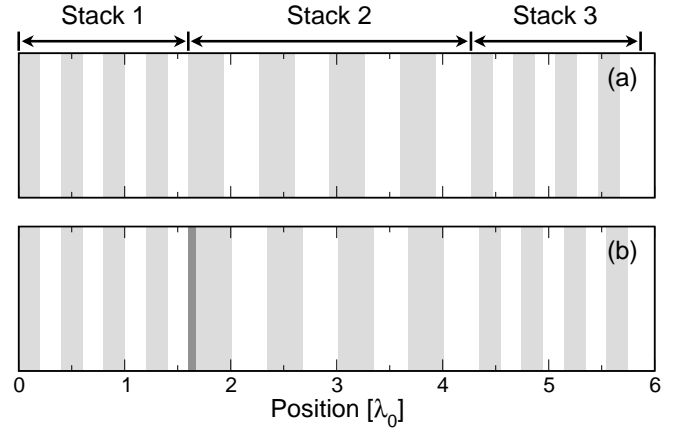
\pict{fig01.eps}{structure}{Schematic of the LH/RH superlattice for
  transmission in (a) linear regime and (b) nonlinear diode
  regime. Layers of LH material are shown in light gray. The
  dielectric defect layer is shown in dark gray.}

We model the LH material as having Lorentzian frequency dependence
in both electric permittivity $\varepsilon_r$ and magnetic
permeability $\mu_r$,
\begin{eqnarray}
  \label{eq:1}
  \varepsilon_r(\omega) &=&
  1+\frac{\omega_{pe}^2}{\omega_{1e}^2-\omega^2-i\gamma_e\omega},\\
  \label{eq:2}
  \mu_r(\omega) &=&
  1+\frac{\omega_{pm}^2}{\omega_{1m}^2-\omega^2-i\gamma_m\omega}.
\end{eqnarray}
Here, $\omega_{pe}, \omega_{pm}$ are the corresponding plasma
frequencies, $\omega_{1e}, \omega_{1m}$ are the resonance frequencies,
and $\gamma_e, \gamma_m$ are absorption parameters. We use
$\omega_{pe}=1.1543\times 10^{11}$~s$^{-1}$, $\omega_{pm}=1.6324\times
10^{11}$~s$^{-1}$, $\omega_{1e}=\omega_{1m}=2\pi\times 5\times
10^6$~s$^{-1}$, and include small losses,
$\gamma_e=2\times\gamma_m=2\pi\times 6\times 10^6$~s$^{-1}$. With
these parameters, the refractive index of the LH material $n\approx-1$
at frequency $f_0=15$~GHz, and the material is left-handed for
frequencies $f<18.5$~GHz and right-handed for $f>26$~GHz. The RH layer
is assumed to be air.

The structure shown in Fig.~\rpict{structure}(a) consists of three
stacks with four LH/RH double layers each. The outer stacks (Stack 1
and Stack 3) have individual layer thicknesses of $\lambda_0/5$,
while the central stack (Stack 2) has individual layer thickness
$\lambda_0/3$; $\lambda_0$ being the free-space wavelength of incoming
radiation of the frequency $f_0$. Figure~\rpict{structure}(b) shows
the same system with a structural defect introduced between Stack
1 and Stack 2. The defect layer has thickness $2\lambda_0/25$.

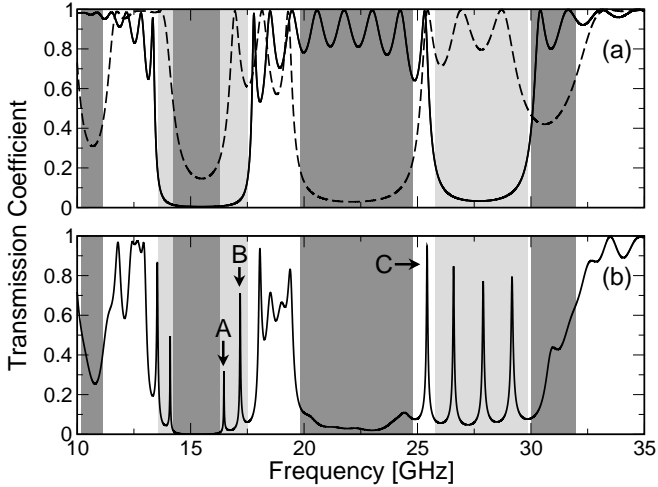
\pict{fig02.eps}{spectrum-linear-superlattice}{Transmission
  spectrum of the structure and its components in the linear regime.
  (a) Stack of eight periods with individual layer
  thickness $\lambda_0/3$ (solid) [Stack 1 and Stack 3 in
  Fig.~\rpict{structure}(a)] and a stack of four periods
  with individual layer thickness $\lambda_0/5$ (dashed) [Stack 2 in
  Fig.~\rpict{structure}(a)]. (b) Combination of these stacks when
  arranged as in Fig.~\rpict{structure}(a). The shading indicates the
  location of the band gaps for Stack~1 (light grey) and Stack~2 (dark
  grey) in the limit of an infinite number of periods.}

First, we study the transmission properties of the individual
components of our superlattice in order to analyze the effect of the
layer thickness on the positions of the band gaps and the transmission
spectrum.  Figure~\rpict{spectrum-linear-superlattice}(a) shows the
transmission spectrum of a structure made of the combination of
Stack~1 and Stack~3 directly attached to each other (solid), as well
as, the transmission spectrum of Stack~2 (dashed).  The spectra show
several transmission bands and gaps, and we clearly see a shift of the
conventional band gaps due to the scaling of the structure.  However,
the band gap associated with vanishing average refractive index
$\left<n\right>=0$ (around $f_0=15$~GHz) remains at the same frequency
and only changes its width~\cite{Li:2003-83901:PRL}.

The transmission spectrum of the combined structure of
Fig.~\rpict{structure}(a) is shown in
Fig.~\rpict{spectrum-linear-superlattice}(b). In the frequency
intervals where the central stack is opaque, the structure has a band
gap, while at frequencies where the outer stacks are opaque but the
central stack is transparent the structure behaves similar to a cavity
resonator and shows the characteristic Fabry-Perot cavity peaks. In
this system cavity-like behavior occurs in the frequency ranges of
13.5--14.3~GHz, 16.3--17.5~GHz, and 25.8--29.9~GHz.

To study the nonlinear transmission of our multi-stack
structure, 
we introduce a defect layer with a linear electric
permittivity $\varepsilon_r=4$, see Fig.~\rpict{structure}(b). For the
TMM calculations, we place 
a delta-function Kerr-type nonlinearity at the left interface of
the defect layer~\cite{Lidorikis:1998-346:PD}
$\varepsilon_r[1+\delta(x-x_0)]$,
  $x_0=1.6\lambda_0$, while in the PSTD calculations the entire
defect layer has a nonlinear electric permittivity
 $\varepsilon_r (t)= 4 + \chi^{(3)}\left|\mathbf{E}(t)\right|^2$,
with the nonlinearity coefficient $\chi^{(3)}=+4$. We
study the difference in transmission when the light is incident
from opposite sides of the structure. In the linear case, the
reciprocity principle holds as expected and the transmission coefficient is
independent of direction of incidence, while a nonlinear
asymmetric structure can show directional dependence of the
transmission indicating the corresponding diode action
effect~\cite{Scalora:1994-2023:JAP,Tocci:1995-2324:APL}.

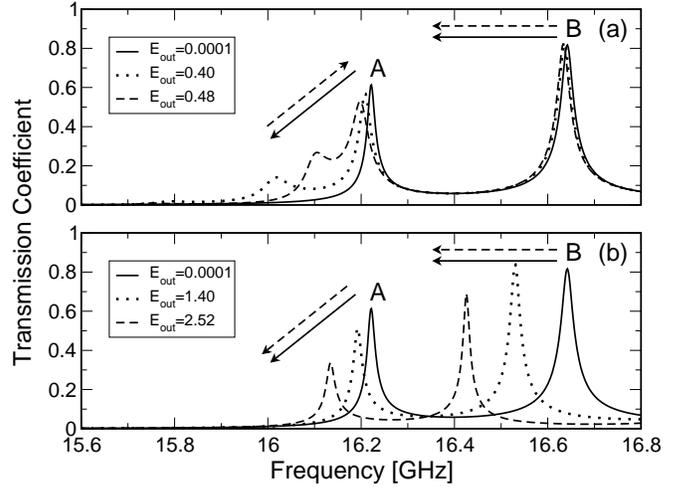
\pict{fig03.eps}{spectrum-nonlinear-superlattice}{Transmission
  spectrum of the structure in Fig.~\rpict{structure}(b) for incidence from the (a) left and (b)
  right, for different output amplitudes, calculated by TMM.
  The arrows indicate the direction in which the peaks shift with
  increased input (solid) and output (dashed) intensity.
}

First, we employ the TMM and study the steady-state results.  In this
situation it is convenient to fix the transmitted (output) field and
calculate backwards to find the incident and reflected fields. In
Fig.~\rpict{spectrum-nonlinear-superlattice} we present a spectral
interval including the first two peaks (peaks A,~B) above the
$\left<n\right>=0$ band gap of Stack 2 for several fixed transmitted
field strengths and different directions of incidence. The inclusion
of the defect shifts all peaks in
Fig.~\rpict{spectrum-linear-superlattice}(b) to slightly lower
frequency in the linear regime.
Due to the nonlinear properties of the defect, the frequencies of the
transmission resonances vary with the local electric field intensity
which in turn varies with the output field.  Since the behavior of
peak B is very similar for the two directions of incidence, the output
field strengths in the figure are chosen to highlight the behavior of
peak A. The frequency shift of peak A occurs at very different output
field strengths depending on the direction of incidence. As the
\emph{input} field is increased, all the transmission peaks shift to
lower frequencies (solid arrows in
Fig.~\rpict{spectrum-nonlinear-superlattice}). The curves are
presented with fixed output field and one notices that all peaks shift
to lower frequencies with increased output field (dashed arrows in
Fig.~\rpict{spectrum-nonlinear-superlattice}), except for peak A in
the case of incidence from the left
[Fig.~\rpict{spectrum-nonlinear-superlattice}(a)], which moves to
higher frequencies. This opposite shift in frequency means that while
generally the output field at the transmission peak increases with
increased input field, for this peak the output field actually
\emph{decreases} as the input field is \emph{increased}.  Referring to
Fig.~\rpict{spectrum-linear-superlattice}(b) one notices that peak A
is located slightly above a band gap of the central stack (Stack 2).
With increased input intensity the nonlinear defect shifts the peak to
lower frequency such that it moves deeper into the band gap and becomes
increasingly suppressed.  Behavior similar to peak A is also found for
the first peak (peak C) above the first Bragg band gap of Stack 2 (not
shown).

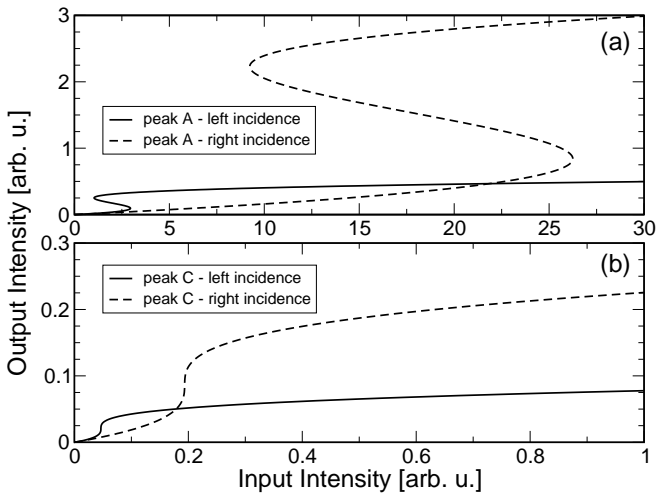
\pict{fig04.eps}{TMM-bistability}{TMM results: Bistability
  curves for $f=16.19$~GHz (a) and $25.17$~GHz (b), at equal relative
  detuning below a linear transmission peaks A and C. The curves show
  incidence from the left (solid) and the right (dashed).  Peak C
  shows bistability for greater but not for smaller detuning than
  shown in (b). Peak A shows bistability even for much smaller
  relative detuning.}

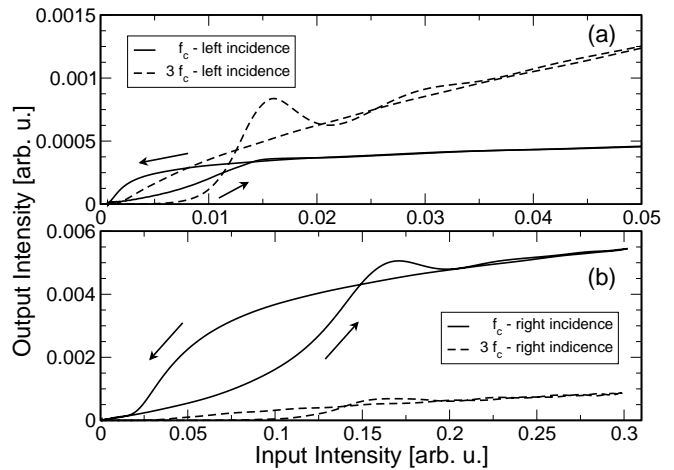
\pict{fig05.eps}{PSTD-hysteresis}{PSTD results: Hysteresis curve
  of the structure for a Gaussian pulse incident from the left (a) and
  the right (b). The incident field
  has carrier frequency $f_c=16.19$~GHz (peak A), width parameter
  $1100/f_c$, and peak amplitude 0.55. The arrows indicate the
  direction of change with time.}

Next, in Fig.~\rpict{TMM-bistability}, we compare the output intensity
as a function of input intensity for incidence from the two opposite
sides at frequencies with equal relative detuning from peak A and peak
C, calculated using the TMM. We find strong nonreciprocity in all
cases.  The transmission in the two directions differs by up to a
factor of 6 for peak A and a factor of 4 for peak C.  Peak A also
shows strong bistability for both directions. The switching intensity
threshold for incidence from the left is almost an order of magnitude
lower than for incidence from the right.
With the given detuning from the linear transmission peak frequency
one finds no bistability for peak C but with stronger detuning
bistability does appear.
Such bistability in an optical diode was previously mentioned in
\cite{Tocci:1995-2324:APL,Chen:2003-1514:CHIL}.

Finally, we simulate the transmission using the PSTD method. This
time-domain method inherently includes the higher harmonics that may
occur due to the nonlinear material. We use a spatial step-size of
$\Delta_x=\lambda_0/75$ and a temporal step-size of
$\Delta_t=\Delta_x/(\pi c)$. The incident field was chosen as a long
Gaussian pulse with peak amplitude $0.55$, carrier frequency
$f_c=16.19$~GHz (peak A) and width parameter $1100/f_c$, to operate in
a regime where the envelope magnitude varies slowly.
Figure~\rpict{PSTD-hysteresis} shows the magnitude of the short-time
Fourier-transform at the carrier frequency and at its third harmonic.
Again, we find that the transmission strongly differs for incidence
from the two sides. Transmission at $f_c$ is initially higher for
incidence from the left but when the incident intensity reaches 0.04
transmission for incidence from the right becomes greater. On the
other hand, the transmission at $3f_c$ is always much greater for the
pulse incident from the left. At $3f_c$ the structure is essentially
transparent, as the LH-material slabs have
$\varepsilon_r\approx\mu_r\approx n\approx 1$ at that frequency. We
find bistability for the carrier frequency, as well as, its third
harmonic with incidence from either direction.  The threshold
switching intensities differ greatly for the two directions.  The
switching intensities for the two frequencies with a particular
direction of incidence are very similar to one another.  These results
agree well with the TMM calculation.

In our PSTD simulations we also encountered
modulational-instability-like behavior in this system with
different threshold intensities for the onset of this instability.
This effect will be subject to further investigation.

In conclusion, we have studied numerically
the linear and nonlinear transmission of a novel type of multilayer
structures composed of two different dielectric materials. We have
shown that asymmetric periodic structures with nonlinear layers
can demonstrate highly nonreciprocal transmission which is a major
characteristic of an optical diode. We have analyzed the
properties of the left-handed diode by employing the
transfer-matrix approach and direct pseudospectral time-domain
numerical simulations, and have shown its bistable behavior.

\end{document}